\documentclass[pre,showkeys,preprintnumbers,amsmath,amssymb,superscriptaddress,onecolumn]{revtex4}
\usepackage[utf8]{inputenc}
\usepackage[T1]{fontenc}
\usepackage{color}
\usepackage[english]{babel}
\usepackage{graphicx}
\usepackage[colorlinks=true, allcolors=blue]{hyperref}

\begin{document}

\title{Pseudo-Darwinian evolution of physical flows in complex networks}

\author{Geoffroy Berthelot}
\affiliation{
Centre de Math\'ematiques Appliqu\'ees, CNRS -- Ecole Polytechnique, IP Paris, 91128 Palaiseau, France}
\affiliation{
Research Laboratory for Interdisciplinary Studies (RELAIS), 75012 Paris, France}
\affiliation{
Institut National du Sport, de l'Expertise et de la Performance (INSEP), 75012 Paris, France}

\author{Liubov Tupikina}
\affiliation{Center for Research and Interdisciplinarity (CRI), Universit\'e de Paris -- INSERM (U1284), 75004 Paris, France}
\affiliation{Bell Labs Nokia, France}

\author{Min-Yeong Kang}
\affiliation{
Laboratoire de Physique de la Mati\`{e}re Condens\'{e}e (UMR 7643), CNRS -- Ecole Polytechnique, IP Paris, 91128 Palaiseau, France}

\author{Bernard Sapoval}
\affiliation{
Laboratoire de Physique de la Mati\`{e}re Condens\'{e}e (UMR 7643), CNRS -- Ecole Polytechnique, IP Paris, 91128 Palaiseau, France}

\author{Denis~S.~Grebenkov}
 \email{denis.grebenkov@polytechnique.edu}
\affiliation{
Laboratoire de Physique de la Mati\`{e}re Condens\'{e}e (UMR 7643), CNRS -- Ecole Polytechnique, IP Paris, 91128 Palaiseau, France}

\date{\today}

\begin{abstract}
The evolution of complex transport networks is investigated under
three strategies of link removal: random, intentional attack and
``Pseudo-Darwinian'' strategy.  At each evolution step and regarding the
selected strategy, one removes either a randomly chosen link, or the
link carrying the strongest flux, or the link with the weakest flux,
respectively.  We study how the network structure and the total flux
between randomly chosen source and drain nodes evolve.  We discover a
universal power-law decrease of the total flux, followed by an abrupt
transport collapse.  The time of collapse is shown to be determined by
the average number of links per node in the initial network,
highlighting the importance of this network property for ensuring safe
and robust transport against random failures, intentional attacks and
maintenance cost optimizations.
\end{abstract}


\maketitle

\section*{Introduction}

Transport in complex systems can describe a variety of natural and
human-engineered processes including biological
\cite{Serov2015,Greb2005}, societal and technological ones
\cite{Newman2004,Lambiotte}. Common examples include blood vessel
network and the lung airway tree \cite{mauroy2004optimal} that deliver
blood and oxygen molecules, respectively; braided streams, consisting
in a network of water channels, that occur in rivers and in glaciated
landscapes when the discharge of water cannot transport its load or
when sediment is deposited on the floor of the channel
\cite{Connor2018,Rinaldo2014}; transportation networks for passengers
\cite{zhang2018review,mattsson2015vulnerability,von2012tale}; social
networks, in which the social and experience flow is progressively
formed between individuals over time. These empirical networks are
often scale-free and characterized by a degree distribution that
follows a power law $P(k) \sim k^{-\gamma}$ with an exponent $\gamma$
often in a range between $2$ and $3$ or a truncated power law
\cite{Barabasi,HavlinStanley}.  The morphological organization of
transport in static scale-free networks was investigated from various
perspectives \cite{Lopez2004,Lopez2007,ResisIpaper,Nicolaides}.  It is
known that transport through scale-free networks and their
functionality in general are vulnerable to the intentional attack to a
few vertices with high degree, but remain very robust to random
failures \cite{Goh2003,Valente04}.  While the percolation by deleting
{\it nodes} has been extensively studied
\cite{Nicolaides,Havlin,RossoSapoval}, the effect of progressive {\it
link} removal on transport in scale-free networks is not well
understood. In this letter we aim at investigating how the temporal
evolution of local connectivities can lead to emergence of ordered
structures in scale-free networks under different strategies of link
removal. While we will focus on physical transport systems, where a
flux is an electric current or the quantity of transferred materials
or molecules over time, the obtained results are of much broader scope
and reveal the fundamental principles of structural evolution of
general transport networks.

\section*{Methods}

We construct a random scale-free network on an $n \times n$ square
lattice.  Its links are generated by using the uncorrelated
configuration model \cite{Satoras} with a given degree exponent
$\gamma$.  We consider the resistance $r_{i,j}$ of each link as a
function of the Euclidean distance $d_{i,j}$ between the nodes $i$ and
$j$: $r_{i,j} =d_{i,j}^{\beta}$, with an exponent $\beta$.  In an
electric or hydraulic circuit, the resistance of a wire or a tube is
proportional to its length, and $\beta=1$.  In turn, most former
studies on transport in resistor networks supposed constant link
resistance, i.e., $\beta=0$.  In each random realization of the
resistor network with prescribed exponents $\gamma$ and $\beta$, we
select randomly a source and a drain nodes, at which the potential is
fixed to be $1$ and $0$, respectively.  We ensure the distance (in
terms of the number of nodes) between the source and drain is $\ge
4$. The system of linear Kirchhoff\textquoteright s equations
\cite{Redner} for the potential on other nodes is solved numerically
using a custom routine in Matlab.  Then the distributions of nodes
potentials and currents in links are obtained.  Such a point-to-point
transport was shown to be self-organized into two tree-like
structures, one emerging from the source and the other converging to
the drain \cite{ResisIpaper}.  These trees merge into a large cluster
of the remaining nodes that is found to be quasi-equipotential and
thus presents almost no resistance to transport.

We consider three dynamics of network evolution, in which links are
progressively removed according to one of the following strategies: at
each step, one removes (i) randomly chosen link, (ii) the link with
the strongest flux, and (iii) the link with the weakest flux. These
three evolution strategies are meant to model respectively (i)
progressive failures in a system in random (unrelated) places (e.g.,
due to material aging); (ii) intentional attacks on the network by
removing the most relevant links; and (iii) a kind of progressive
optimization of the system by removing least used elements. While the
first two evolution strategies have been earlier studied (but mainly
for nodes removal) \cite{Gallos2005}, we are not aware of former works
on the latter network evolution that we call ``Pseudo-Darwinian strategy'' of
network percolation.  Such a strategy is often employed in nature,
e.g., as the mechanism of capillary network remodeling during
morphogenesis \cite{LeNoble2004}.

At each evolution step, we first solve the system of
Kirchhoff\textquoteright s equations to calculate fluxes in all links,
and then we remove a link according to the selected strategy. After
link removal, we also remove ``dead-ends'' (i.e., nodes that have a
single link), thus imposing that any existing node after an
evolutionary step has at least two links.  We keep track of the
\emph{total flux} $Q$, i.e., the flux that enters into the network
from the source node.  As we investigate the evolution of the network
after successive link removals, it is natural to associate the number
of evolution steps with ``time'' $t$, with $0$ being the initial time,
before the evolution takes place.  We denote $Q_0$, $L_0$ and $N_0 =
n^2$ as the total flux, the number of links, and the number of nodes
of the initial network, respectively.
The evolution ends when at least one of the following conditions is
met: (i) no path exists between the source and the drain (i.e. the
source and drain are disconnected), (ii) the source or the drain is
removed from the network, (iii) a portion of the network -- a subgraph
containing more than one node -- is disconnected from the rest of the
network containing the source and drain.  We call this moment as
``time of collapse'' $t_c$: as one of the previous conditions is met,
transport is no longer maintained through the whole network.

\begin{figure*}
\centering
\includegraphics[width=1\textwidth]{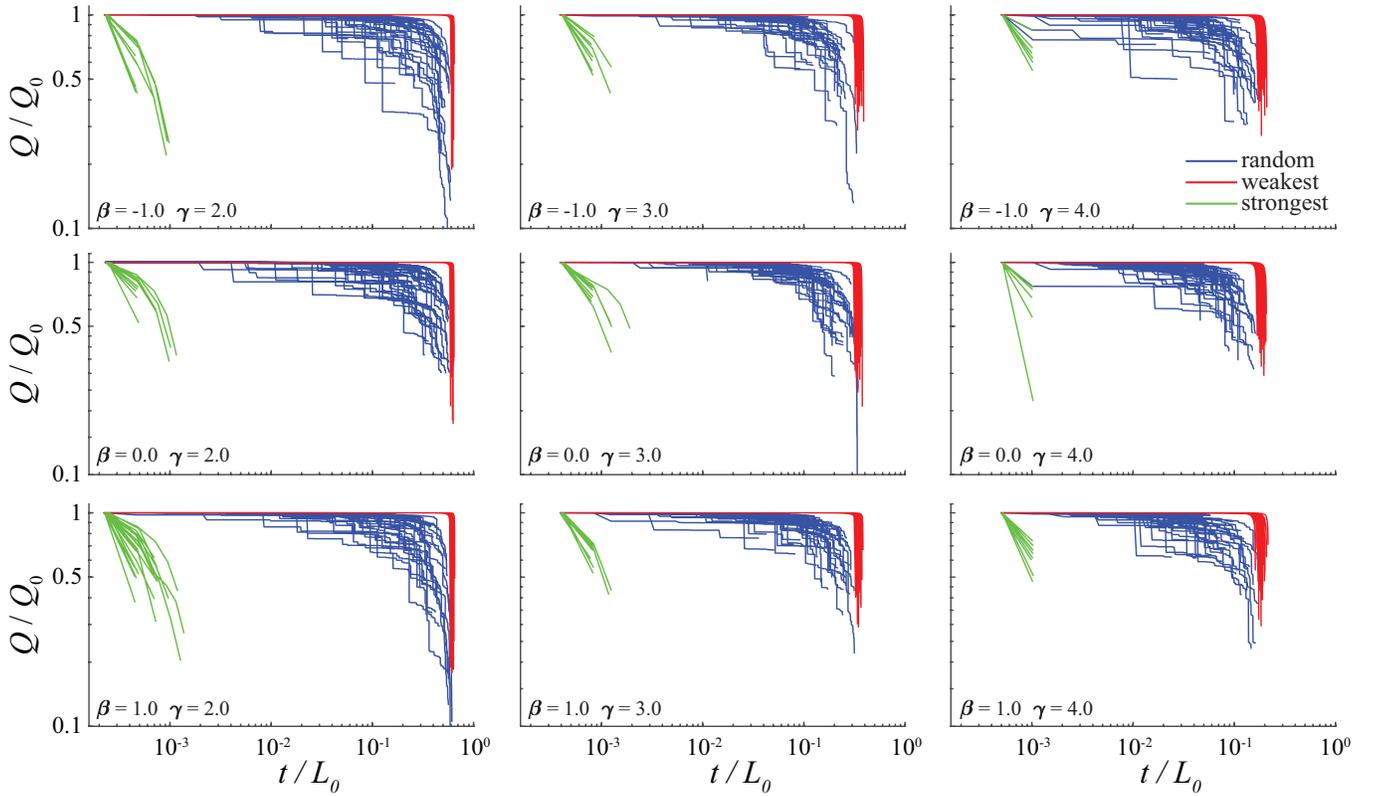}
\caption{
Evolution of the total flux $Q$ (rescaled by $Q_0$) for a scale-free
network with $N_0 = 40\times 40$ nodes for nine combinations of
parameters $\gamma$ and $\beta$: $\gamma = \{2,3,4\}$ (left, middle
and right columns) and $\beta = \{-1,0,1\}$ (top, middle and bottom
rows).  In each plot, 60 curves correspond to 60 random realizations
of the initial network.}
\label{fig:flux}
\end{figure*}

\section*{Results}

\begin{figure}
\centering
\includegraphics[width=0.45\textwidth]{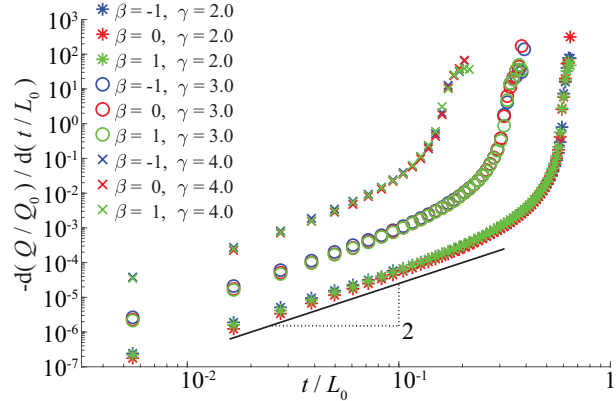} 
\caption{
Time derivative of the total flux for a scale-free network with $N_0 =
40\times40$ nodes and different values of $\gamma$ and $\beta$ for the
pseudo-Darwinian strategy only.  The slope is identical for all values of
$\gamma$ and $\beta$, suggesting a universal phenomenon. The
derivative here is calculated as an average from 60 realizations.}
\label{fig:derivative}
\end{figure}

\begin{figure}
\centering
\includegraphics[width=0.45\textwidth]{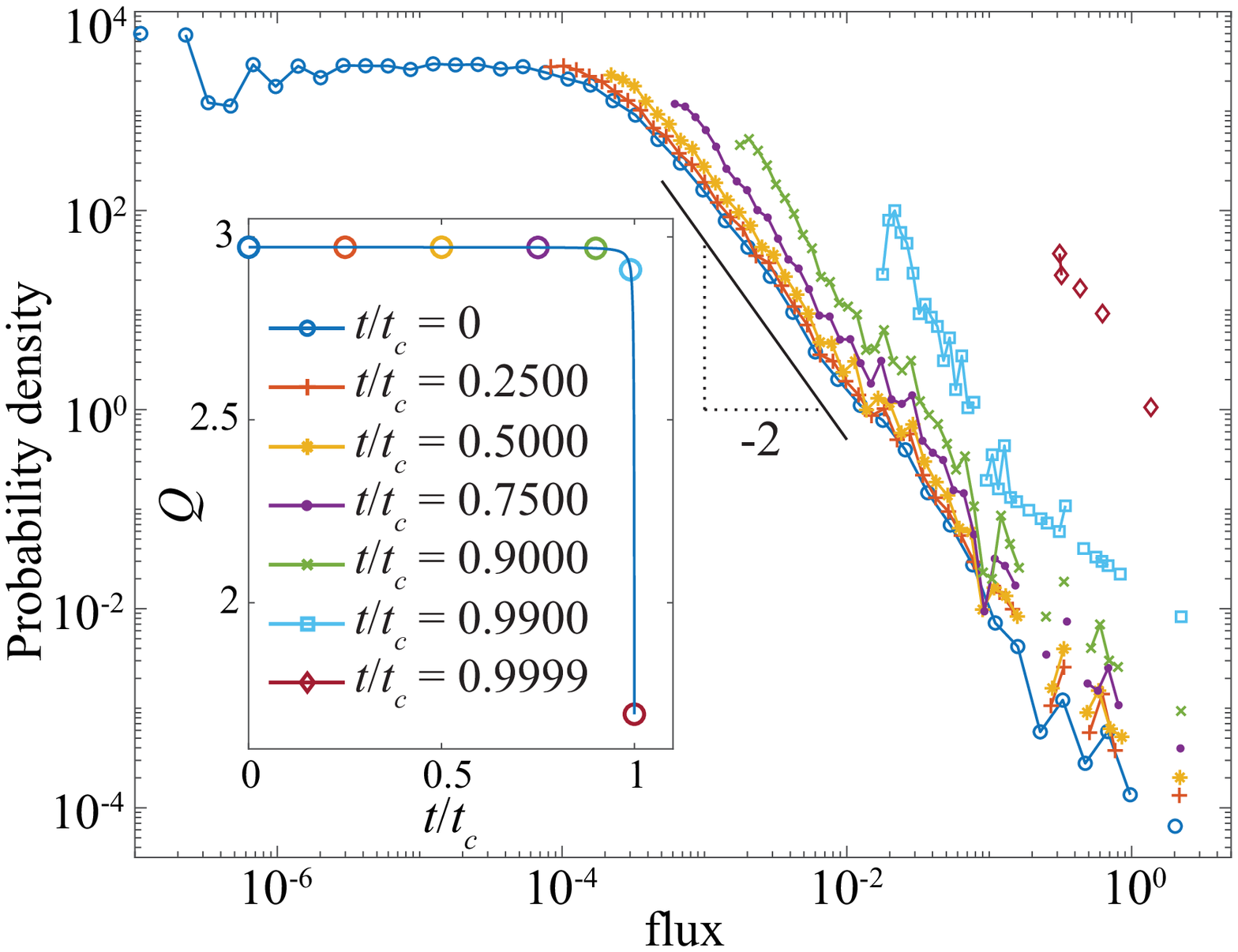} 
\caption{
The flux probability density at selected moments of $t/t_c$, for a
network with $N_0 = 100 \times 100$ nodes, $\gamma = 2.5$, $\beta =
1$, and $L_0 = 21040$ initial links.  Corresponding total fluxes are
shown in the inset.}
\label{fig:fluxevol}
\end{figure}

\begin{figure}
\centering
\includegraphics[width=0.45\textwidth]{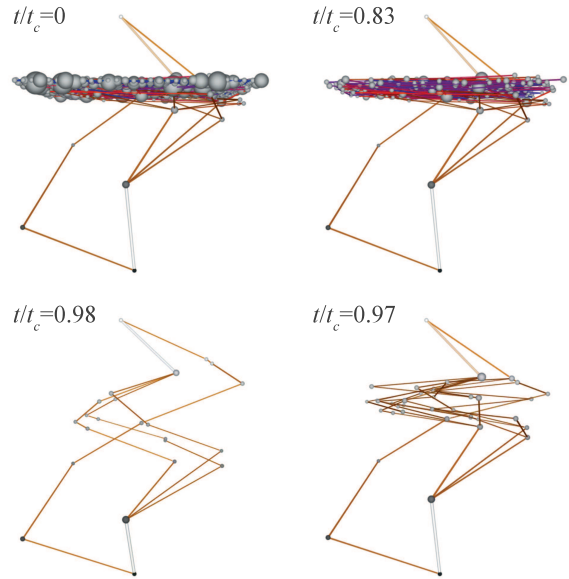} 
\caption{
Visualization of evolution of a scale-free network (with $N_0
=30\times30$ nodes, $\gamma=2.5$ and $\beta=1$; A video is available
online).  Each node of the network is shown by a ball whose radius is
proportional to the square root of its connectivity.  The planar
coordinates of the balls are the positions of the corresponding nodes
on the square lattice, whereas the height $Z$ represents the potential
at the node by a linear relation $Z=V$.  Each link brightness is
proportional to the magnitude of its current (in addition, blue colors
are used for very small currents).  The quasi-equipotential cluster is
qualitatively identified as a large ensemble of nodes almost at the
same potential.  Four panels represent the network at four moments,
$t/t_c$, during evolution (clock-wise direction): 0 (initial state),
0.83, 0.97, and 0.98.  Along the evolution process, the weakest links
are removed successively.  Note that links in the cluster are removed
during the majority of evolution period.}
\label{fig:Animation}
\end{figure}

We first explore a scale-free resistor network with $N_0 =40\times 40$
nodes and parameters $\gamma\in\{2, 3, 4\}$ and $\beta\in\{-1,0,1\}$.
Figure~\ref{fig:flux} shows how the total flux $Q$, rescaled by its
initial value $Q_0$, evolves with time $t$ for different networks
($\gamma$, $\beta$) and strategies.  Expectedly, the progressive
removal of the strongest links leads to a very rapid transportation
collapse, whereas this decline is the slowest for the pseudo-Darwinian
strategy. To better understand the effect of this strategy, we also
calculate the derivative of the total flux $Q$ with respect to time
(Fig.~\ref{fig:derivative}). Two regimes are observed: a slow, power
law decay, followed by an abrupt decay. In the first regime, we find
the universal scaling exponent $2$ of the derivative of the total flux
for all $\gamma$ and $\beta$ values, implying
\begin{equation}
Q(t)/Q_0 \simeq 1 - C (t/t_c)^3 \qquad (t \ll t_c),
\end{equation}
where $C$ is a nonuniversal constant. The distribution of fluxes in
links along the evolution process (Fig.~\ref{fig:fluxevol}) shows that
the removal of the weakest link at each evolution step does not affect
the distribution of large fluxes for the most period of evolution. For
this period, the total flux remains almost the same. Figure
\ref{fig:Animation} helps to understand this robustness: the removal
occurs mostly in the links in the quasi-equipotential cluster where
connections are abundant while the links with large currents are kept.

\begin{figure}
\centering
\includegraphics[width=0.99\textwidth]{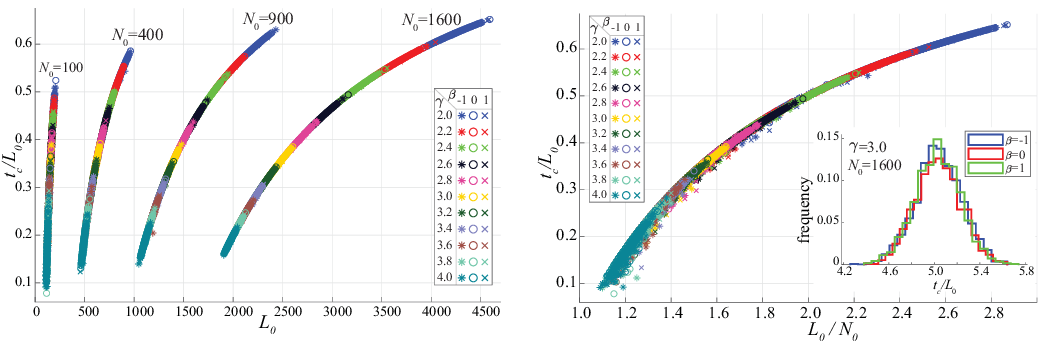}
\caption{
Rescaled time of collapse $t_c / L_0$ versus the initial number of
links $L_0$ {\bf (left)} or versus the average initial number of links
per node $L_0 / N_0$ {\bf (right)} for 4 different network sizes
($N_0=100$, $N_0=400$, $N_0=900$, $N_0=1600$), 11 values of $\gamma$
(evenly spaced from 2 to 4) and 3 values of $\beta$
($\beta=\{-1,0,1\}$).  Each colored marker represents the time of
collapse obtained from 200 realizations for each triplet of $\gamma$
(specified by color), $\beta$ (specified by marker) and network size
$N_0$.  The parameter $\beta$ has no significant effect on $t_c$, as
confirmed in the inset, where three colored lines show the empirical
distributions of $t_c$ for three different values of $\beta$ (obtained
from 2000 simulations).  }
\label{fig:tc_Lzero}
\end{figure}

The time of collapse $t_c$ is related to the degree exponent $\gamma$
in all strategies: networks with lower $\gamma$ values produce higher
$t_c$ (Fig.~\ref{fig:flux}).  This highlights the role of connectivity
in resisting against random or targeted attacks.  This effect is
illustrated for the pseudo-Darwinian evolution by plotting the time of
collapse rescaled by $L_0$, $t_c / L_0$, versus the initial number of
links $L_0$ for various network sizes (Fig.~\ref{fig:tc_Lzero}(left)).
Further rescaling the horizontal axis $L_0$ by the initial number of
nodes, $N_0$, results in a collapse of all curves into a single master
curve that determines $t_c/L_0$ as a function of the initial average
degree, $L_0/N_0$, independently of the network structure ($\gamma$,
$\beta$) and size $N_0$ (Fig.~\ref{fig:tc_Lzero}(right)).  Thus we
conclude that higher average degree in an initial network helps in
delaying the time $t_c$ of transport collapse.

\section*{Discussion}

In our setup, the total flux describes a network functionality and its
ability to transport or distribute a current. Thus, the sudden
collapse observed in all three strategies means that the network
eventually fails to transport. The onset of this collapse is affected
by the evolution strategy, in particular, the pseudo-Darwinian strategy leads
to a nearly ``optimized'' network, i.e. a structure that carries most
of the flux with minimal number of useless links (i.e. links that
carry a negligible part of the flux). This can also be seen as an
``economical'' structure, where the transport function is kept with
the least size of components and, hence, minimized maintenance cost
\cite{Murray1926}. However, going to an optimal system makes it
fragile and dangerous because a small disturbance can lead to a sudden
collapse. Therefore, for robustness of the network, a safety margin
from the critical point $t_c$ should be considered, as known for human
bronchial systems \cite{mauroy2004optimal}.

When reaching $t_c$, the ``optimized'' structure becomes a chain
connecting pre-selected source and drain nodes.  While this structure
is formally optimal, it is impractical for applications due to its
``optimality'' for the particular choice of the source and the drain.
A much more challenging and practically relevant question would be the
construction of the optimal structure for all (or for most) pairs of
source and drain nodes.  This problem will be investigated in a
subsequent work.

This letter focused on the evolution of both transport and structural
properties of scale-free networks under progressive link removal.
Such transport-driven dynamics further extend common models of network
evolutions \cite{Dorogovtsev02,Dorogovtsev08,Garlaschelli07}.  In
particular, our analysis helps to investigate the precursors of the
transitions for links percolation applied to resistor networks
\cite{Rodriguez} or epidemic spreading \cite{Sander2003}.  More
generally, this study can serve as a basis for illustrating generic
evolution dynamics of complex networks governed by its transport
properties. Different works already made the analogy between
Kirchhoff\textquoteright s laws and the conservation of mass equation,
with applications to vehicular flow \cite{guarnaccia2010analysis},
fractures in materials \cite{batrouni1998fracture}, neuronal circuitry
in the brain \cite{destexhe2012neurons}.  For instance, a direct
analogy with random walks on graphs opens exciting perspectives for
understanding diffusive transport and first-passage processes on
evolving networks \cite{Condamin07,Tupikina19}.  Our approach is also
expected to stimulate further studies of link-based percolation in
other networks such as in protein networks, where links can be lost
with time (corresponding to the loss of some proteins'
functionalities), while proteins, the nodes of a network, remain
present
\cite{kovacs2019network}.

\section*{Acknowledgements}

We thank Jean-Fran\c{c}ois Colonna for the creation of the video
showing the evolution of the network (Fig. \ref{fig:Animation}).
L.~T. acknowledges a partial finantial support via the CRI Research
Fellowship thanks to the Bettencourt Schueller Foundation long term
partnership.  D.~S.~G. acknowledges a partial financial support from
the Alexander von Humboldt Foundation through a Bessel Research Award.

\end{document}